\documentclass[aps,showpacs,twocolumn,superscriptaddress]{revtex4}
\usepackage{amssymb}
\usepackage{epsfig}
\usepackage{graphicx,amsmath}
\begin{document}

\title{Fermion condensation: a strange idea successfully explaining
behavior of numerous objects in Nature}
\author{V.R. Shaginyan}\email{vrshag@thd.pnpi.spb.ru}
\affiliation{Petersburg Nuclear Physics Institute, RAS, Gatchina,
188300, Russia}
\author{M.Ya. Amusia}\affiliation{Racah Institute
of Physics, Hebrew University, Jerusalem 91904, Israel}
\author{K.G. Popov}\affiliation{Komi Science Center, Ural Division,
RAS, Syktyvkar, 167982, Russia}

\begin{abstract}
Strongly correlated Fermi systems are among the most intriguing,
best experimentally studied and fundamental systems in physics.
These are, however, in defiance of theoretical understanding. The
ideas based on the concepts like Kondo lattice and involving
quantum and thermal fluctuations at a quantum critical point have
been used to explain the unusual physics. Alas, being suggested to
describe one property, these approaches fail to explain the others.
This means a real crisis in theory suggesting that there is a
hidden fundamental law of nature, which remains to be recognized. A
theory of fermion condensation quantum phase transition, preserving
the extended quasiparticles paradigm and intimately related to
unlimited growth of the effective mass as a function of
temperature, magnetic field etc, is capable to resolve the problem.
We discuss the construction of the theory and show that it delivers
theoretical explanations of vast majority of experimental results
in strongly correlated systems such as heavy-fermion metals and
quasi-two-dimensional Fermi systems. Our analysis is placed in the
context of recent salient experimental results. Our calculations of
the non-Fermi liquid behavior, of the scales and thermodynamic and
transport properties are in good agreement with the heat capacity,
magnetization, longitudinal magnetoresistance and magnetic entropy
obtained in remarkable measurements on the heavy fermion metal $\rm
YbRh_2Si_2$. Using two-dimensional $\rm ^3He$ as an example, we
demonstrate that the main universal features of its experimental
temperature $T$ - density $x$ phase diagram resemble those of the
heavy-fermion metals. We propose a simple expression for the
effective mass, describing all diverse experimental facts on the
$\rm ^3He$ in unified manner and demonstrating that the universal
behavior of the effective mass coincides with that observed in
heavy fermion metals.\\

\end{abstract}

 \maketitle

\section{Introduction}

Strongly correlated Fermi systems represented by heavy fermion (HF)
metals, high-temperature superconductors and quasi-two-dimensional
$^3$He are among the most intriguing, best experimentally studied
and fundamental systems in physics \cite{A1}. This is also a field
never far from applications in synthesis of novel materials for
cryogenics, rare earth magnets and applied superconductivity. Their
behavior is so unusual that the traditional Landau quasiparticles
paradigm does not apply to it. The paradigm states that the
properties is determined by quasiparticles whose dispersion is
characterized by the effective mass $M^*$ which is independent of
temperature $T$, the number density $x$, magnetic field $B$ and
other external parameters. The above systems are, however, in
defiance of theoretical understanding. The ideas based on the
concepts (like Kondo lattice, see e.g. Ref. \cite{loh}) involving
quantum and thermal fluctuations at a quantum critical point (QCP)
have been used to explain the unusual physics of these systems
known as non-Fermi liquid (NFL) behavior \cite{A1,loh,si,sach}.
Alas, being suggested to describe one property, these approaches
fail to explain the others. This means a real crisis in theory
suggesting that there is a hidden fundamental law of nature, which
remains to be recognized \cite{col}. It is widely believed that
utterly new concepts are required to describe the underlying
physics. There is a fundamental question: how many concepts do we
need to describe the above physical mechanisms? This cannot be
answered on purely experimental or theoretical grounds. Rather, we
have to use both of them.

Usual arguments that quasiparticles in strongly correlated Fermi
liquids "get heavy and die" at a quantum critical point commonly
employ the well-known formula basing on assumptions that the
$z$-factor (the quasiparticle weight in the single-particle state)
vanishes at the points of second-order phase transitions
\cite{col1}. However, it has been shown that this scenario is
problematic \cite{khodz}. A concept of fermion condensation quantum
phase transition (FCQPT) preserving quasiparticles and intimately
related to the unlimited growth of $M^*$, had been suggested
\cite{khs,ams,zph,volovik}. Studies show that it is capable to
deliver an adequate theoretical explanation of vast majority of
experimental results in different HF metals \cite{obz,khodb,shag3}.
In contrast to the Landau paradigm based on the assumption that
$M^*$ is a constant, in FCQPT approach $M^*$ strongly depends on
$T$, $x$, $B$ etc. Therefore, in accord with numerous experimental
facts the extended quasiparticles paradigm is to be introduced. The
main point here is that the well-defined quasiparticles determine
as before the thermodynamic and transport properties of strongly
correlated Fermi-systems, while $M^*$ becomes a function of $T$,
$x$, $B$, and the dependence of the effective mass on $T$, $x$, $B$
gives rise to the non-Fermi liquid (NFL) behavior
\cite{obz,khodb,shag3,zph,ckz,plaq,jetpl}.

In this review report we discuss the construction of a theory,
based on the above FCQPT approach and its application to the
analysis of wide variety of experimentally observed phenomena in
microscopically different strongly correlated Fermi systems like
heavy-fermion metals and quasi-two-dimensional $^3$He.  We analyze
the NFL behavior of strongly correlated Fermi systems and show that
this is generated by the dependence of the effective mass on
temperature, number density and magnetic field at FCQPT. We
demonstrate that the NFL behavior observed in the transport and
thermodynamic properties of HF metals can be described in terms of
the scaling behavior of the normalized effective mass. This allows
us to construct the scaled thermodynamic and transport properties
extracted from experimental facts in wide range of the variation of
scaled variable. We show that "peculiar points" of the normalized
effective mass give rise to the energy scales observed in the
thermodynamic and transport properties of HF metals. Our
calculations of the thermodynamic and transport properties are in
good agreement with the heat capacity, magnetization, longitudinal
magnetoresistance and magnetic entropy obtained in remarkable
measurements on the heavy fermion metal $\rm YbRh_2Si_2$
\cite{steg1,oes,steg,geg}.

\section{Fermion condensation quantum phase transition}

We start with visualizing the main properties of FCQPT. To this
end, consider the density functional theory for superconductors
(SCDFT) \cite{gross}. SCDFT states that at fixed temperature $T$
the thermodynamic potential $\Phi$ is a universal functional of the
number density $n({\bf r})$ and the anomalous density (or the order
parameter) $\kappa({\bf r},{\bf r}_1)$ and provides a variational
principle to determine the densities \cite{gross}. At the
superconducting transition temperature $T_c$ a superconducting
state undergoes the second order phase transition. Our goal now is
to construct a quantum phase transition which evolves from the
superconducting one. In that case, the superconducting state takes
place at $T=0$ while at finite temperatures there is a normal
state. This means that in this state the anomalous density is
finite while the superconducting gap vanishes. For the sake of
simplicity, we consider a homogeneous Fermi (electron) system.

Let us assume that the coupling constant $\lambda$ of the pairing
interaction vanishes, $\lambda\to0$, making vanish the
superconducting gap at any finite temperature. In that case,
$T_c\to0$ and the superconducting state takes place at $T=0$ while
at finite temperatures there is a normal state. This means that at
$T=0$ the anomalous density is finite while the superconducting gap
is infinitely small \cite{zph,obz,shag1}. For the sake of
simplicity, we consider a homogeneous Fermi (electron) system
\cite{obz}. Then, the thermodynamic potential reduces to the ground
state energy $E$ which turns out to be a functional of the
occupation number $n({\bf p})$ since $\kappa=\sqrt{n(1-n)}$
\cite{dft,gross,yakov,plaq,jetpl}. Upon minimizing $E$ with respect
to $n({\bf p})$, we obtain
\begin{equation}\label{FCM}
\frac{\delta E}{\delta n({\bf p})}=\varepsilon({\bf
p})=\mu,\end{equation} where $\mu$ is the chemical potential. It is
seen from Eq. \eqref{FCM} that instead of the Fermi step, we have
$0<n(p)<1$ in certain range of momenta $p_i\leq p\leq p_f$ with
$\kappa$ is finite in this range. Thus, the step-like Fermi filling
inevitably undergoes restructuring and formes the fermion condensate
(FC) as soon as Eq. \eqref{FCM} possesses not-trivial solutions at
some point $x=x_c$ when $p_i=p_f=p_F$ \cite{khs,obz,khodb}. Here
$p_F$ is the Fermi momentum and $x =p_F^3/3\pi^2$.

At any small but finite temperature the anomalous density $\kappa$
(or the order parameter) decays and this state undergoes the first
order phase transition and converts into a normal state
characterized by the thermodynamic potential $\Phi_0$. At $T\to0$,
the entropy $S=-\partial \Phi_0/\partial T$ of the normal state is
given by the well-known relation \cite{land}
\begin{equation}
S_0=-2\int[n({\bf p}) \ln (n({\bf p}))+(1-n({\bf p})\ln (1-n({\bf
p}))]\frac{d{\bf p}}{(2\pi) ^3},\label{SN}
\end{equation}
which follows from combinatorial reasoning. Since the entropy of the
superconducting ground state is zero, it follows from Eq. \eqref{SN}
that the entropy is discontinuous at the phase transition point,
with its discontinuity $\Delta S=S_0$. The latent heat $q$ of
transition from the asymmetrical to the symmetrical phase is
$q=T_cS_0=0$ since $T_c=0$. Because of the stability condition at
the point of the first order phase transition, we have
$\Phi_0[n({\bf p})]=\Phi[\kappa({\bf p})]$. Obviously the condition
is satisfied since $q=0$.

At $T=0$, a quantum phase transition is driven by a nonthermal
control parameter, e.g. the number density $x$. To clarify the role
of $x$, consider the effective mass $M^*$ which is related to the
bare electron mass $m$ by the well-known Landau equation
\cite{land} which is also valid when $M^*$ strongly depends on $B$,
$T$ or $x$ \cite{plaq}
\begin{equation}\label{LANDM}
\frac{1}{M^*}=\frac{1}{m}+\int \frac{{\bf p}_F{\bf p_1}}{p_F^3}
F({\bf p_F},{\bf p}_1)\frac{\partial n(p_1,T)}{\partial p_1}
\frac{d{\bf p}_1}{(2\pi)^3}.
\end{equation}
Here we omit the spin indices for simplicity, $n({\bf p},T)$ is
quasiparticle occupation number, and $F$ is the Landau amplitude. At
$T=0$, Eq. \eqref{LANDM} reads \cite{pfit,pfit1}
\begin{equation}\label{MM*}
\frac{M^*}{m}=\frac{1}{1-N_0F^1(x)/3}.\end{equation} Here $N_0$ is
the density of states of free electron gas and $F^1(x)$ is the
$p$-wave component of Landau interaction amplitude $F$. When at
some quantum critical point (QCP) $x=x_c$, $F^1(x)$ achieves
certain threshold value, the denominator in Eq. \eqref{MM*} tends
to zero so that the effective mass diverges at $T=0$
\cite{pfit,pfit1}. It follows from Eq. \eqref{MM*} that beyond the
QCP $x=x_c$, the effective mass becomes negative. To avoid unstable
and physically meaningless state with a negative effective mass,
the system must undergo a quantum phase transition at QCP $x=x_c$
defined by Eq. \eqref{FCM} and which is FCQPT
\cite{khs,ams,obz,khodb}.

\begin{figure} [! ht]
\begin{center}
\vspace*{-0.5cm}
\includegraphics [width=0.49\textwidth]{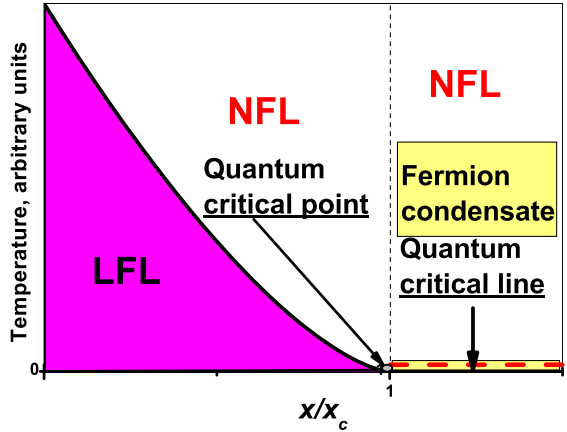}
\end{center}
\vspace*{-0.8cm} \caption{Schematic phase diagram of the system
driven to the FC state. The number density $x$ is taken as the
control parameter and depicted as $x/x_c$. The quantum critical
point (QCP), $x/x_c=1$, of FCQPT is shown by the arrow. At
$x/x_c<1$ and sufficiently low temperatures, the system is in the
Landau Fermi liquid (LFL) state as shown by the shadow area. At
$T=0$ and beyond QCP, $x/x_c>1$, the system is at the quantum
critical line depicted by the dash line and shown by the vertical
arrow. The critical line is characterized by the FC state with
finite superconducting order parameter $\kappa$. At any finite
temperature, the order parameter $\kappa$ is destroyed and the
entropy becomes discontinuous at $T_c=0$, the system undergoes the
first order phase transition and exhibits the NFL behavior at
$T>0$.}\label{fig1}
\end{figure}

Schematic phase diagram of the system which is driven to FC by
variation of $x$ is reported in Fig. \ref{fig1}.  Upon approaching
the critical density $x_c$ the system remains in LFL region at
sufficiently low temperatures \cite{khodb,obz}, that is shown by the
shadow area. At QCP $x_c$ shown by the arrow in Fig. \ref{fig1}, the
system demonstrates the NFL behavior down to the lowest
temperatures. Beyond QCP at finite temperatures the behavior is
remaining the NFL one and is determined by the
temperature-independent entropy $S_0$ \cite{yakov}. In that case at
$T\to 0$, the system is approaching a quantum critical line (shown
by the vertical arrow and the dashed line in Fig. \ref{fig1}) rather
than a quantum critical point. Upon reaching the quantum critical
line from the above at $T\to0$ the system undergoes the first order
quantum phase transition, which is FCQPT taking place at $T_c=0$.

At $T>0$ the NFL state above the critical line, see Fig.
\ref{fig1}, is strongly degenerated, therefore it is captured by
the other states such as superconducting (for example, by the
superconducting state in $\rm CeCoIn_5$ \cite{bi,shag1,shag2}) or
by AF state (e.g. AF one in $\rm YbRh_2Si_2$ \cite{geg1,geg,plaq})
lifting the degeneracy. The application of magnetic field
$B>B_{c0}$ restores the LFL behavior, where $B_{c0}$ is a critical
magnetic field, such that at $B>B_{c0}$ the system is driven
towards its Landau Fermi liquid (LFL) regime \cite{geg1,geg,shag2}.
In some cases, for example in HF metal $\rm CeRu_2Si_2$,
$B_{c0}=0$, see e.g. \cite{takah}, while in $\rm YbRh_2Si_2$,
$B_{c0}\simeq 0.06$ T \cite{geg1,geg}. In our simple model $B_{c0}$
is taken as a parameter.

\section{Scaling behavior of the effective mass}

\begin{figure}[!ht]
\begin{center}
\includegraphics [width=0.44\textwidth]{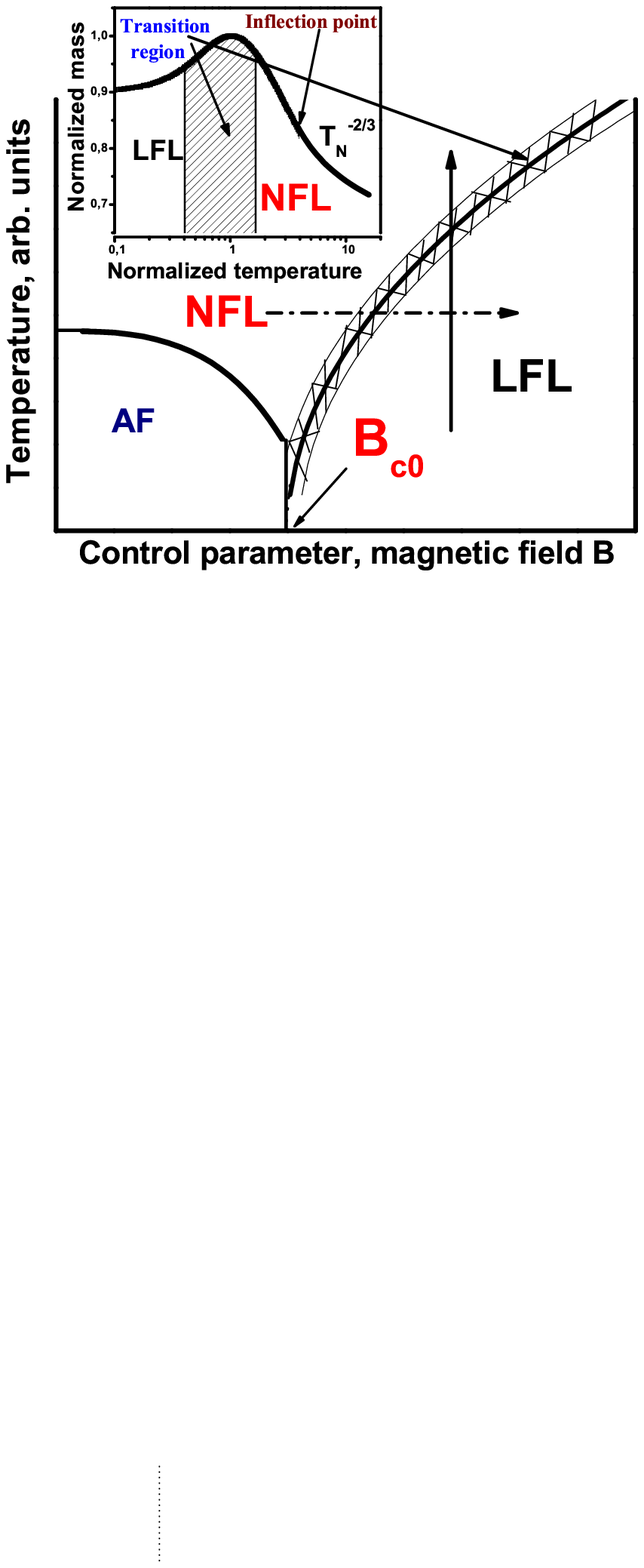}
\vspace*{-11.8cm}
\end{center}
\caption{ Schematic phase diagram of the HF metal $\rm YbRh_2Si_2$.
$\rm {AF}$ denotes antiferromagnetic state. At $T=0$, $B_{c0}$ is
magnetic field at which the effective mass diverges and the AF
state vanishes. At $B>B_{c0}$ the system is in its paramagnetic
state. The vertical arrow shows the transition from the LFL regime
to the NFL one at fixed $B$ along $T$ with $M^*$ depending on $T$.
The dash-dot horizontal arrow illustrates the system moving from
NFL regime to LFL one along $B$ at fixed $T$. The inset shows a
schematic plot of the scaling behavior of the normalized effective
mass versus the normalized temperature. Transition regime, where
$M^*_N$ reaches its maximum value $M^*_M$ at $T=T_M$, is shown by
the hatched area both in the main panel and in the inset. The
arrows mark the position of inflection point in $M^*_N$ and the
transition region.}\label{PHD}
\end{figure}

Schematic phase diagram of the HF metal $\rm YbRh_2Si_2$ is
reported in Fig. \ref{PHD}. Magnetic field $B$ is taken as the
control parameter. The FC state and the region lying at $x/x_c\geq
1$, see Fig. \ref{fig1}, can be captured by the superconducting,
ferromagnetic, antiferromagnetic (AF) etc. states lifting the
degeneracy \cite{obz,khodb}. Since we consider the HF metal $\rm
YbRh_2Si_2$ the AF state takes place \cite{geg1,geg} as shown in
Fig. \ref{PHD}. As seen from Fig. \ref{PHD}, at elevated
temperatures and fixed magnetic field the NFL regime occurs, while
rising $B$ again drives the system from NFL region to LFL one.
Below we consider the transition region when at rising $B$ the
system moves from NFL regime to LFL one along the dash-dot
horizontal arrow, and at elevated $T$ it moves from LFL regime to
NFL one along the solid vertical arrow.

To explore a scaling behavior of $M^*$, we write the quasiparticle
distribution function as $n_1({\bf p})=n({\bf p},T)-n({\bf p})$,
with $n({\bf p})$ is the step function, and Eq. \eqref{LANDM} then
becomes
\begin{equation}
\frac{1}{M^*(T)}=\frac{1}{M^*}+\int\frac{{\bf p}_F{\bf
p_1}}{p_F^3}F({\bf p_F},{\bf p}_1)\frac{\partial n_1(p_1,T)}
{\partial p_1}\frac{d{\bf p}_1}{(2\pi) ^3}. \label{LF1}
\end{equation}
At QCP the effective mass $M^*$ diverges and Eq. \eqref{LF1} becomes
homogeneous determining $M^*$ as a function of temperature
\begin{equation}M^*(T)\propto T^{-2/3},\label{LTT}
\end{equation}
while the system exhibits the NFL behavior \cite{ckz,obz}. If the
system is located before QCP, $M^*$ is finite, at low temperatures
the system demonstrates the LFL behavior that is $M^*(T)\simeq
M^*+a_1T^2$, with $a_1$ is a constant, see the inset to Fig.
\ref{PHD}. Obviously, the LFL behavior takes place when the second
term on the right hand side of Eq. \eqref{LF1} is small in
comparison with the first one. Then, at rising temperatures the
system enters the transition regime: $M^*$ grows, reaching its
maximum $M^*_M$ at $T=T_M$, with subsequent diminishing. Near
temperatures $T\geq T_M$ the last "traces" of LFL regime disappear,
the second term starts to dominate, and again Eq. \eqref{LF1}
becomes homogeneous, and the NFL behavior restores, manifesting
itself in decreasing $M^*$ as $T^{-2/3}$, see Eq. \eqref{LTT}. When
the system is near QCP, it turns out that the solution of Eq.
\eqref{LF1} $M^*(T)$ can be well approximated by a simple universal
interpolating function \cite{obz,ckz,shag2}. The interpolation
occurs between the LFL ($M^*\simeq M^*+a_1T^2$) and NFL
($M^*\propto T^{-2/3}$) regimes thus describing the above crossover
\cite{ckz,obz}. Introducing the dimensionless variable
$y=T_N=T/T_M$, we obtain the desired expression
\begin{equation}M^*_N(y)\approx c_0\frac{1+c_1y^2}{1+c_2y^{8/3}}.
\label{UN2}
\end{equation}
Here $M^*_N=M^*/M^*_M$ is the normalized effective mass,
$c_0=(1+c_2)/(1+c_1)$, $c_1$ and $c_2$ are fitting parameters,
parameterizing the Landau amplitude.

The inset to Fig. \ref{PHD} demonstrates the scaling behavior of the
normalized effective mass $M^*_N=M^*/M^*_M$ versus normalized
temperature $T_N=T/T_M$, where $M^*_M$ is the maximum value that
$M^*$ reaches at $T=T_M$. At $T\ll T_M$ the LFL regime takes place.
At $T\gg T_M$ the $T^{-2/3}$ regime takes place. This is marked as
NFL one since the effective mass depends strongly on temperature.
The temperature region $T\simeq T_M$ signifies the transition
between the LFL regime with almost constant effective mass and NFL
behavior, given by $T^{-2/3}$ dependence. Thus temperatures $T\sim
T_M$ can be regarded as the transition region between LFL and NFL
regimes. The transition temperatures are not really a phase
transition. These necessarily are broad, very much depending on the
criteria for determination of the point of such a transition, as it
is seen from the inset to Fig. \ref{PHD}. As usually, the transition
temperature is extracted from the temperature dependence of charge
transport, for example, from the resistivity $\rho(T)=\rho_0+AT^2$
with $\rho_0$ is the residual resistivity and $A$ is the LFL
coefficient. The crossover takes place at temperatures where the
resistance starts to deviate from the LFL $T^2$ behavior. Obviously,
the measure of the deviation from the LFL $T^2$ behavior cannot be
defined unambiguously. Therefore, different measures produce
different results.

It is possible to transport Eq. \eqref{LF1} to the case of the
application of magnetic fields \cite{ckz,obz,shag2}.  The
application of magnetic field restores the LFL behavior so that
$M^*_M$ depends on $B$ as
\begin{equation}\label{LFLB}
    M^*_M\propto (B-B_{c0})^{-2/3},
\end{equation} while
\begin{equation}\label{LFLT}T_M\propto \mu_B(B-B_{c0}),\end{equation}
where $\mu_B$ is the Bohr magneton \cite{ckz,shag2,obz}. Employing
Eqs. \eqref{LFLB} and \eqref{LFLT} to calculate $M^*_M$ and $T_M$,
we conclude that Eq. \eqref{UN2} is valid to describe the normalized
effective mass in external fixed magnetic fields with
$y=T/(B-B_{c0})$. On the other hand, Eq. \eqref{UN2} is valid when
the applied magnetic field becomes a variable, while temperature is
fixed $T=T_f$. In that case, as seen from Eqs. \eqref{LTT},
\eqref{UN2} and\eqref{LFLB}, it is convenient to rewrite both the
variable as $y=(B-B_{c0})/T_f$, and Eq. \eqref{LFLT} as
\begin{equation}\label{LFLf}\mu_B(B_M-B_{c0})\propto T_f.\end{equation}

It follows from Eq. \eqref{UN2} that in contrast to the Landau
paradigm of quasiparticles the effective mass strongly depends on
$T$ and $B$. As we will see it is this dependence that forms the
NFL behavior. It follows also from Eq. \eqref{UN2} that a scaling
behavior of $M^*$ near QCP is determined by the absence of
appropriate external physical scales to measure the effective mass
and temperature. At fixed magnetic fields, the characteristic
scales of temperature and of the function $M^*(T,B)$ are defined by
both $T_M$ and $M^*_M$ respectively. At fixed temperatures, the
characteristic scales are $(B_M-B_{c0})$ and $M^*_M$. It follows
from Eqs. \eqref{LFLB} and \eqref{LFLT} that at fixed magnetic
fields, $T_M\to0$, and $M^*_M\to\infty$, and the width of the
transition region shrinks to zero as $B\to B_{c0}$ when these are
measured in the external scales. In the same way, it follows from
Eqs. \eqref{LTT} and \eqref{LFLf} that at fixed temperatures,
$(B_M-B_{c0})\to0$, and $M^*_M\to\infty$, and the width of the
transition region shrinks to zero as $T_f\to0$. Thus, the
application of the external scales obscure the scaling behavior of
the effective mass and of the thermodynamic and transport
properties.

A few remarks are in order here. As we shall see, magnetic field
dependencies of the effective mass or of other observable like the
longitudinal magnetoresistance do not have "peculiar points" like
maximum. The normalization are to be performed in the other points
like the inflection point at $T=T_{inf}$ (or at $B=B_{inf}$) shown
in the inset to Fig. \ref{PHD} by the arrow. Such a normalization
is possible since it is established on the internal scales,
$T_{inf}\propto T_M\propto(B-B_{c0})$.

\section{NFL behavior of the HF metal $\rm \bf YbRh_2Si_2$}

\begin{figure} [! ht]
\begin{center}
\vspace*{-0.5cm}
\includegraphics [width=0.49\textwidth]{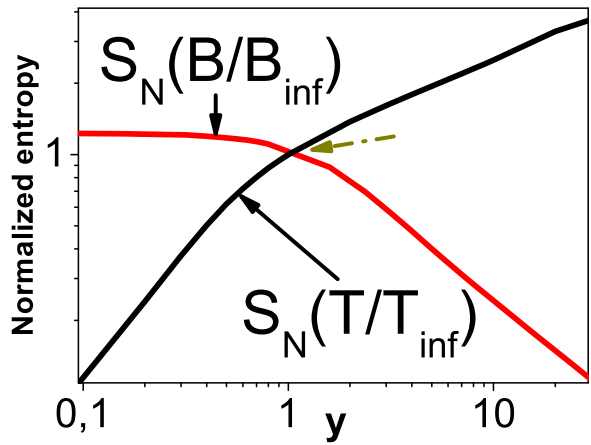}
\end{center}
\vspace*{-1.2cm} \caption{The normalized entropy $S_N(B/B_{inf})$
versus $y=B/B_{inf}$ and the normalized entropy $S_N(T/T_{inf})$
versus $y=T/T_{inf}$ calculated at fixed temperature and magnetic
field, correspondingly, are represented by the solid lines and
shown by the arrows. The inflection point is depicted by the
dash-dot arrow.}\label{STB}
\end{figure}

In what follows, we compute the effective mass and employ Eq.
\eqref{UN2} for estimations of considered values. To compute the
effective mass $M^*(T,B)$, we solve Eq. \eqref{LF1} with special
form of Landau interaction amplitude, see Refs. \cite{ckz,obz} for
details. Choice of the amplitude is dictated by the fact that the
system has to be at QCP, which means that first two $p$-derivatives
of the single-particle spectrum $\varepsilon({\bf p})$ should equal
zero. Since first derivative is proportional to the reciprocal
quasiparticle effective mass $1/M^*$, its zero just signifies QCP
of FCQPT. Zero of the second derivative means that the spectrum
$\varepsilon({\bf p})$ has an inflection point at $p_F$ rather than
a maximum. Thus, the lowest term of the Taylor expansion of
$\varepsilon({\bf p})$ is proportional to $(p-p_F)^3$ \cite{ckz}.
After solution of Eq. \eqref{LF1}, the obtained spectrum had been
used to calculate the entropy $S(B,T)$, which, in turn, had been
recalculated to the effective mass $M^*(T,B)$ by virtue of
well-known LFL relation $M^*(T,B)=S(T,B)/T$. Our calculations of
the normalized entropy as a function of the normalized magnetic
field $B/B_{inf}=y$ and as a function of the normalized temperature
$y=T/T_{inf}$ are reported in Fig. \ref{STB}. Here $T_{inf}$ and
$B_{inf}$ are the corresponding inflection points in function $S$.
We normalize the entropy by its value at the inflection point
$S_N(y)=S(y)/S(1)$. As seen form Fig. \ref{STB}, our calculations
corroborate the scaling behavior of the normalized entropy, that is
the curves at different temperatures and magnetic fields merge into
single one in terms of the variable $y$. The inflection point
$T_{inf}$ in $S(T)$ makes $M^*(T,B)$ have its maximum as a function
of $T$, while $M^*(T,B)$ versus $B$ has no maximum. We note that
our calculations of the entropy confirm the validity of Eq.
\eqref{UN2} and the scaling behavior of the normalized effective
mass.

\subsection{Heat capacity}

Exciting measurements of $C/T\propto M^*$ on samples of the new
generation of $\rm YbRh_2Si_2$ in different magnetic fields $B$ up
to 1.5 T \cite{oes} allow us to identify the scaling behavior of the
effective mass $M^*$ and observe the different regimes of $M^*$
behavior such as the LFL regime, transition region from LFL to NFL
regimes, and the NFL regime itself. A maximum structure in
$C/T\propto M^*_M$ at $T_M$ appears under the application of
magnetic field $B$ and $T_M$ shifts to higher $T$ as $B$ is
increased. The value of $C/T=\gamma_0$ is saturated towards lower
temperatures decreasing at elevated magnetic field, where $\gamma_0$
is the Sommerfeld coefficient \cite{oes}.

The transition region corresponds to the temperatures where the
vertical arrow in the main panel of Fig. \ref{PHD} crosses the
hatched area. The width of the region, being proportional to
$T_M\propto (B-B_{c0})$ shrinks,  $T_M$ moves to zero temperature
and $\gamma_0\propto M^*$ increases as $B\to B_{c0}$. These
observations are in accord with the facts \cite{oes}.

\begin{figure} [! ht]
\begin{center}
\vspace*{-0.2cm}
\includegraphics [width=0.49\textwidth]{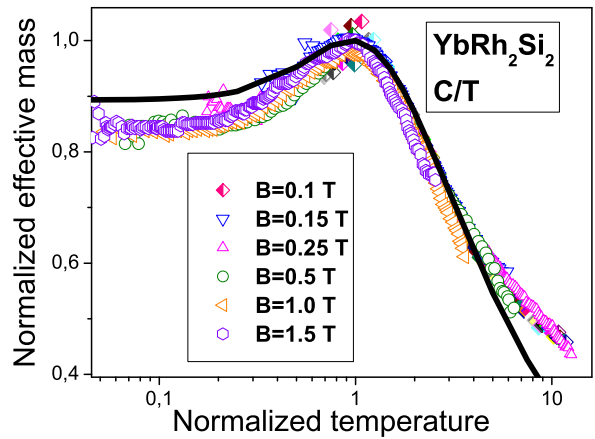}
\end{center}
\vspace*{-0.3cm} \caption{The normalized effective mass $M^*_N$
extracted from the measurements of the specific heat $C/T$ on $\rm
YbRh_2Si_2$ in magnetic fields $B$ \cite{oes} listed in the legend.
Our calculations are depicted by the solid curve tracing the scaling
behavior of $M^*_N$.}\label{fig2}
\end{figure}

To obtain the normalized effective mass $M^*_N$, the maximum
structure in $C/T$ was used to normalize $C/T$, and $T$ was
normalized by $T_M$. In Fig. \ref{fig2} $M^*_N$ as a function of
normalized temperature $T_N$ is shown by geometrical figures, our
calculations are shown by the solid line. Figure \ref{fig2} reveals
the scaling behavior of the normalized experimental curves - the
scaled curves at different magnetic fields $B$ merge into a single
one in terms of the normalized variable $y=T/T_M$. As seen, the
normalized mass $M^*_N$ extracted from the measurements is not a
constant, as would be for  LFL. The two regimes (the LFL regime and
NFL one) separated by the transition region, as depicted by the
hatched area in the inset to Fig. \ref{PHD}, are clearly seen in
Fig. \ref{fig2} illuminating good agreement between the theory and
the facts. It is worthy of note that the normalization procedure
allows us to construct the scaled function $C/T$ extracted from the
facts in wide range variation of the normalized temperature.
Indeed, it integrates measurements of $C/T$ taken at the
application of different magnetic fields into unique function which
demonstrates the scaling behavior over three decades in normalized
temperature as seen from Fig. \ref{fig2}.

\subsection{Magnetization}

Consider now the magnetization $M$ as a function of magnetic field
$B$ at fixed temperature $T=T_f$
\begin{equation}\label{CHIB}
M(B,T)=\int_0^B \chi(b,T)db,
\end{equation}
where the magnetic susceptibility $\chi$ is given by \cite{land}
\begin{equation}\label{CHI}
\chi(B,T)=\frac{\beta M^*(B,T)}{1+F_0^a}.
\end{equation}
Here, $\beta$ is a constant and $F_0^a$ is the Landau amplitude
related to the exchange interaction \cite{land}. In the case of
strongly correlated systems $F_0^a\geq -0.9$ \cite{pfit,pfit1}.
Therefore, as seen from Eq. \eqref{CHI}, due to the normalization
the coefficients $\beta$ and $(1+F_0^a)$ drops out from the result,
and $\chi\propto M^*$.

\begin{figure} [! ht]
\begin{center}
\vspace*{-0.5cm}
\includegraphics [width=0.49\textwidth]{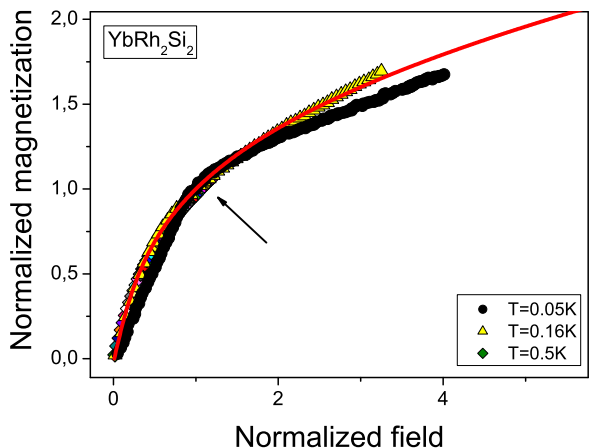}
\end{center}
\vspace*{-0.8cm} \caption{The field dependencies of the normalized
magnetization $M$ collected at different temperatures shown at
right bottom corner are extracted from measurements collected on
$\rm{ YbRu_2Si_2}$ \cite{steg}. The kink (shown by the arrow) is
clearly seen at the normalized field $B_N=B/B_k\simeq 1$. The solid
curve represents our calculations.}\label{fig3}
\end{figure}

One might suppose that $F_0^a$ can strongly depend on $B$. This is
not the case, since the Kadowaki-Woods ratio is conserved
\cite{kadw,geg1,kwz,natphys}, $A(B)/\gamma_0^2(B)\propto
A(B)/\chi^2(B)\propto const$, we have $\gamma_0\propto M^*\propto
\chi$. Here $A$ is the coefficient in the $T^2$ dependence of
resistivity $\rho$.

Our calculations show that the magnetization exhibits a kink at
some magnetic field $B=B_k$. The experimental magnetization
demonstrates the same behavior \cite{steg}. We use $B_k$ and
$M(B_k)$ to normalize $B$ and $M$ respectively. The normalized
magnetization $M(B)/M(B_k)$ extracted from facts \cite{steg}
depicted by the geometrical figures and calculated magnetization
shown by the solid line are reported in Fig. \ref{fig3}. As seen,
the scaled data at different $T_f$ merge into a single one in terms
of the normalized variable $y=B/T_k$. It is also seen, that these
exhibit energy scales separated by kink at the normalized magnetic
field $B_N=B/B_k=1$. The kink is a crossover point from the fast to
slow growth of $M$ at rising magnetic field. It is seen from Fig.
\ref{fig3}, that our calculations are in good agreement with the
facts, and all the data exhibit the kink (shown by the arrow) at
$B_N\simeq 1$ taking place as soon as the system enters the
transition region corresponding to the magnetic fields where the
horizontal dash-dot arrow in the main panel of Fig. \ref{PHD}
crosses the hatched area. Indeed, as seen from Fig. \ref{fig3}, at
lower magnetic fields $M$ is a linear function of $B$ since $M^*$
is approximately independent of $B$. Then, it follows from Eqs.
\eqref{UN2} and \eqref{LFLB} that at elevated magnetic fields $M^*$
becomes a diminishing function of $B$ and generates the kink in
$M(B)$ separating the energy scales discovered in Refs.
\cite{steg1,steg}. Then, as seen from Eq. \eqref{LFLf} the magnetic
field $B_k$ at which the kink appears, $B_k\simeq B_M\propto T_f$,
shifts to lower $B$ as $T_f$ is decreased. This observation is in
accord with the facts \cite{steg1,steg}.

\subsection{Longitudinal magnetoresistance}

\begin{figure} [! ht]
\begin{center}
\vspace*{-0.5cm}
\includegraphics [width=0.49\textwidth]{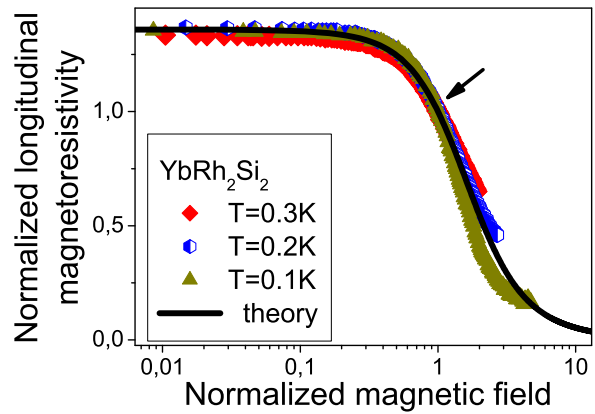}
\end{center}
\vspace*{-0.8cm} \caption{Magnetic field dependence of the
normalized magnetoresistance $\rho_N$ versus normalized magnetic
field. $\rho_N$ was extracted from LMR of $\rm YbRh_2Si_2$ at
different temperatures \cite{steg1,steg} listed in the legend. The
inflection point is shown by the arrow, and the solid line
represents our calculations.}\label{fig4}
\end{figure}

Consider a longitudinal magnetoresistance (LMR)
$\rho(T,B)=\rho_0+A(T,B)T^2$ as a function of $B$ at fixed $T=T_f$.
In that case, the classical contribution to LMR due to orbital
motion of carriers induced by the Lorentz force is small, while the
Kadowaki-Woods relation \cite{kadw,geg1,kwz,natphys},
$K=A/\gamma_0^2\propto A/\chi^2=const$, allows us to employ $M^*$
to construct the coefficient $A$ \cite{pla3}, since
$\gamma_0\propto\chi\propto M^*$. As a result,
$\rho(T,B)-\rho_0\propto(M^*)^2$. Fig. \ref{fig4} reports the
normalized magnetoresistance
\begin{equation}\label{rn}
\rho_N(y)=\frac{\rho(y)-\rho_0}{\rho_{inf}}\propto (M_N^*(y))^2
\end{equation}
versus normalized magnetic field $y=B/B_{inf}$ at different
temperatures, shown in the legend. Here $\rho_{inf}$ and $B_{inf}$
are LMR and magnetic field respectively taken at the inflection
point marked by the arrow in Fig. \ref{fig4}. Both theoretical
(shown by the solid line) and experimental (marked by the
geometrical figures) curves have been normalized by their inflection
points, which also reveals the scaling behavior - the scaled curves
at different temperatures merge into single one as a function of the
variable $y$ and show the scaling behavior over three decades in the
normalized magnetic field. The transition region at which LMR starts
to decrease is shown in the inset to Fig. \ref{PHD} by the hatched
area. Obviously, as seen from Eq. \eqref{LFLf}, the width of the
transition region being proportional to $B_M\simeq B_{inf}\propto
T_f$ decreases as the temperature $T_f$ is lowered. In the same way,
the inflection point of LMR, generated by the inflection point of
$M^*$ shown in the inset to Fig. \ref{PHD} by the arrow, shifts to
lower $B$ as $T_f$ is decreased. All these observations are in
excellent agreement with the facts \cite{steg1,steg}.

\begin{figure} [! ht]
\begin{center}
\vspace*{-0.6cm}
\includegraphics [width=0.47\textwidth]{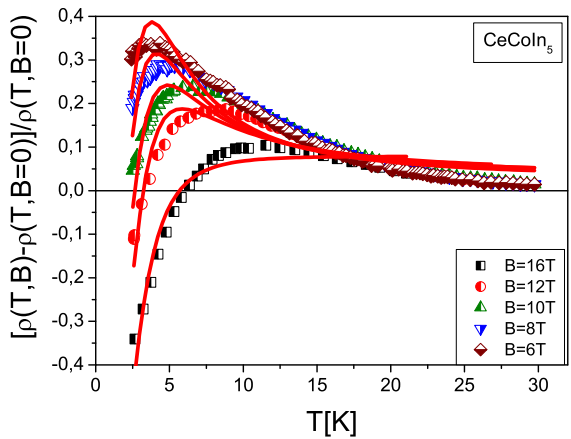}
\end {center}\vspace{-1.0cm}
\caption {MR versus temperature $T$ as a function of magnetic field
$B$. The experimental data on MR were collected on $\rm{ CeCoIn_5}$
at fixed magnetic field $B$ \cite{pag} shown in the right bottom
corner of the Figure. The solid lines represent our
calculations.}\label{MRT}
\end{figure}

It is instructive to demonstrate that the same effective mass
employed to calculate LMR shown in Fig. \ref{fig4} gives good
description of the magnetoresistance (MR) collected in measurements
on $\rm{ CeCoIn_5}$. Figure \ref{MRT} shows the calculated MR
versus temperature as a function of magnetic field $B$ together
with the experimental points from Ref. \cite{pag}. We note that
both the classical contribution to MR due to orbital motion of
carriers induced by the Lorentz force and $\rho_0$ were omitted. As
seen from Fig. \ref{MRT}, our description of experiment is pretty
good \cite{plamr}.

\subsection{Magnetic entropy}

The evolution of the derivative of magnetic entropy $dS(B,T)/dB$ as
a function of magnetic field $B$ at fixed temperature $T_f$ is of
great importance since it allows us to study the scaling behavior of
the derivative of the effective mass $TdM^*(B,T)/dB\propto
dS(B,T)/dB$. While the scaling properties of the effective mass
$M^*(B,T)$ can be analyzed via LMR, see Fig. \ref{fig4}.

\begin{figure} [! ht]
\begin{center}
\includegraphics [width=0.47\textwidth]{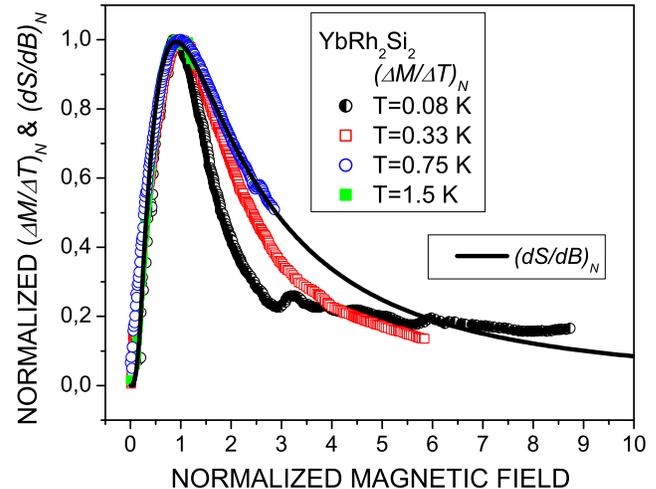}
\vspace*{-0.6cm}
\end{center}
\caption{Normalized magnetization difference divided by temperature
increment $(\Delta M/\Delta T)_N$ versus normalized magnetic field
at fixed temperatures listed in the legend is extracted from the
facts collected  on $\rm YbRh_2Si_2$ \cite{geg}. Our calculation of
the normalized derivative $(dS/dB)_N\simeq (\Delta M/\Delta T)_N$
versus normalized magnetic field is shown by the solid line.}
\label{fig5}
\end{figure}

As seen from from Eqs. \eqref{UN2} and \eqref{LFLf}, at $y\leq 1$
the derivative $-dM_N(y)/dy\propto y$ with
$y=(B-B_{c0})/(B_{inf}-B_{c0})\propto (B-B_{c0})/T_f$. We note that
the effective mass as a function of $B$ does not have the maximum.
At elevated $y$ the derivative $-dM_N(y)/dy$ possesses a maximum at
the inflection point and then becomes a diminishing function of $y$.
Upon using the variable $y=(B-B_{c0})/T_f$, we conclude that at
decreasing temperatures, the leading edge of the function
$-dS/dB\propto -TdM^*/dB$ becomes steeper and its maximum at
$(B_{inf}-B_{c0})\propto T_f$ is higher. These observations are in
quantitative agreement with striking measurements of the
magnetization difference divided by temperature increment, $-\Delta
M/\Delta T$, as a function of magnetic field at fixed temperatures
$T_f$ collected on $\rm YbRh_2Si_2$ \cite{geg}. We note that
according to the well-know thermodynamic equality $dM/dT=dS/dB$, and
$\Delta M/\Delta T\simeq dS/dB$. To carry out a quantitative
analysis of the scaling behavior of $-dM^*(B,T)/dB$, we calculate
the normalized entropy $S$ shown in Fig. \ref{STB} as a function of
$B/B_{inf}$ at fixed temperature $T_f$. Fig. \ref{fig5} reports the
normalized $(dS/dB)_N$ as a function of the normalized magnetic
field. The scaled function $(dS/dB)_N$ is obtained by normalizing
$(-dS/dB)$ by its maximum taking place at $B_M$, and the field $B$
is scaled by $B_M$. The measurements of $-\Delta M/\Delta T$ are
normalized in the same way and depicted in Fig. \ref{fig5} as
$(\Delta M/\Delta T)_N$ versus normalized field. It is seen from
Fig. \ref{fig5} that our calculations shown by the solid line are in
good agreement with the facts and the scaled functions $(\Delta
M/\Delta T)_N$ extracted from the facts show the scaling behavior in
wide range variation of the normalized magnetic field $B/B_M$.

\subsection{Energy scales and $T-B$
phase diagram for $\rm \bf YbRh_2Si_2$}

\begin{figure} [! ht]
\begin{center}
\vspace*{-0.6cm}
\includegraphics [width=0.49\textwidth]{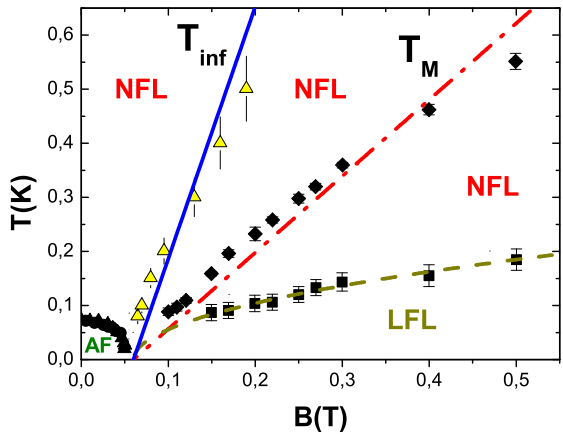}
\end{center}
\vspace*{-0.8cm} \caption{Temperature versus magnetic field $T-B$
phase diagram for $\rm YbRh_2Si_2$. Solid circles represent the
boundary between AF and NFL states. The solid squares  denote the
boundary of the NFL and LFL regime \cite{steg1,steg,geg1} shown by
the dotted line which is approximated by $\sqrt{B-B_{c0}}$
\cite{obz}. Diamonds mark the maximums $T_M$ of $C/T$ \cite{oes}
shown in Fig. \ref{fig2}. The dash-dot line is approximated by
$T_M\propto a(B-B_{c0})$, $a$ is a fitting parameter, see Eq.
\eqref{LFLT}. Triangles along the solid line denote $T_{inf}$ in LMR
\cite{steg1,steg} sown in Fig. \ref{fig5}, the solid line represents
the function $T_{inf}\propto b(B-B_{c0})$, $b$ is a fitting
parameter, see Eq. \eqref{LFLf}.}\label{fig6}
\end{figure}

Fig. \ref{fig6} reports $T_{inf}$ and $T_M$ versus $B$ depicted by
the solid and dash-dotted lines, respectively. The boundary between
the NFL and LFL regimes is shown by the dashed line, and AF marks
the antiferromagnetic state. The corresponding data are taken from
Ref. \cite{steg1,steg,oes,geg1}. It is seen that our calculations
are in good agreement with the facts \cite{jetpl}. In Fig.
\ref{fig6}, the solid and dash-dotted lines corresponding to the
functions $T_{inf}$ and $T_M$, respectively, represent the
positions of the kinks separating the energy scales in $C$ and $M$
reported in Ref. \cite{steg1,steg}. It is seen that our
calculations are in accord with facts, and we conclude that the
energy scales are reproduced by Eqs. \eqref{LFLT} and \eqref{LFLf}
and related to the peculiar points of the normalized effective mass
$M^*_N$. The points are the inflection point $T_{inf}$ and the
maximum point $T_M$ at which the transition region is located.
These are shown by the arrows in the inset to Fig. \ref{PHD}.

At $B\to B_{c0}$ both $T_{inf}\to 0$ and $T_{M}\to 0$, thus the LFL
and the transition regimes of both $C/T$ and $M$ as well as these of
LMR and the magnetic entropy are shifted to very low temperatures.
Therefore due to experimental difficulties these regimes cannot be
often observed in experiments on HF metals. As it is seen from Figs.
\ref{fig2}, \ref{fig3}, \ref{fig4}, \ref{fig5} and \ref{fig6}, the
normalization allows us to construct the unique scaled thermodynamic
and transport functions extracted from the experimental facts in
wide range of the variation of the scaled variable $y$. As seen from
the mentioned Figures, the constructed normalized thermodynamic and
transport functions show the scaling behavior over three decades in
the normalized variable.

\section{Universal Behavior of Two-Dimensional $\rm \bf^3He$ at
Low Temperatures}

The bulk liquid $\rm ^3He$ is historically the first object, to
which the Landau Fermi-liquid (LFL) theory had been applied
\cite{land}. This substance, being an intrinsically isotropic
Fermi-liquid with negligible spin-orbit interaction is ideal to
test the LFL theory. It is remarkable that the same $^3$He becomes
the first 2D homogeneous Fermi-liquid in which the NFL behavior has
been detected  \cite{he3,he3a,prlhe}. 2D $^3$He has a very
important feature: a change of the number density $x$ of $^3$He
film drives it towards QCP at which the quasiparticle effective
mass $M^*$ diverges \cite{he3,he3a,prlhe}. This peculiarity permits
to plot the experimental temperature-density phase diagram, which
can be directly compared with the theoretical phase diagram shown
in Fig. \ref{fig1}. As a result, 2D $\rm ^3He$ becomes an ideal
system to test a theory describing the NFL behavior. Namely, the
neutral atoms of $\rm ^3He$ are fermions interacting with each
other by Van-der-Vaals forces with strong hardcore repulsion and a
weakly attractive tail. The different character of inter-particle
interaction along with the fact, that the mass of $\rm ^3He$ atom
is 3 orders of magnitude larger than that of an electron, makes
$\rm ^3He$ systems to have drastically different properties than
those of HF metals. Because of this difference nobody can be sure
that the macroscopic physical properties of these systems will be
more or less similar to each other at their QCP.

In this Section we show that despite of very different microscopic
nature of 2D $^3$He and 3D HF metals, their main universal features
at their QCP are the same, being dictated by the extended
quasiparticles paradigm. Our analysis of the experimental
measurements has shown that the behavior of 2D $^3$He is quite
similar to that of HF compounds with various ground state magnetic
properties. Namely, we demonstrate that the main universal features
of $\rm ^3He$ experimental $T$-$x$ phase diagram resemble those in
HF metals and can be well captured utilizing the notion of FCQPT
embracing the extended quasiparticles paradigm and thus deriving
NFL properties of above systems from the paradigm. We also show
that the universal behavior of the effective mass of 2D $\rm ^3He$
coincides with that observed in HF metals.

\subsection{The temperature-number density phase
diagram of 2D $\rm ^3He$}

As we seen in Section I, at QCP $x=x_c$ the effective mass diverges
at $T=0$ and the leading term of this divergence given by Eq.
\eqref{MM*} reads
\begin{equation}\label{zui2}
\frac{M^*(x)}{M}=A+\frac{B}{1-z},\ z=\frac {x}{x_c}.
\end{equation}
Equation (\ref{zui2}) is valid in both 3D and 2D cases, while the
values of factors $A$ and $B$ depend on dimensionality and
inter-particle interaction \cite{obz}. At $x>x_c$ the fermion
condensation takes place. Here we confine ourselves to the case
$x<x_c$.

Equation \eqref{zui2} shows that the maximum value of the effective
mass $M^*_M\propto 1/(1-z)$ and it follows from \eqref{LTT} that
$M^*_M\propto T^{-2/3}$. As a result, we obtain that $T_M$ at which
the effective mass reaches its maximum value $M^*_M\propto
T^{-2/3}$ is given by
\begin{equation}\label{zui4}
T_M\propto(1-z)^{3/2}.
\end{equation}
We note that obtained results are in agreement with numerical
calculations \cite{obz,ckz}.
\begin{figure} [! ht]
\begin{center}
\includegraphics [width=0.4\textwidth]{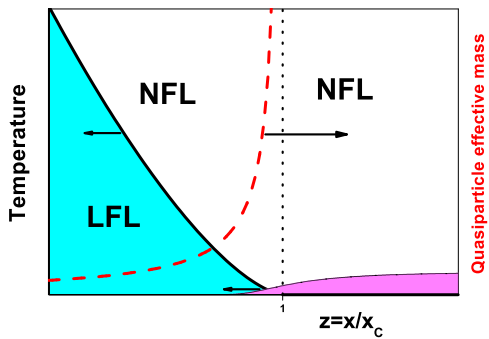}
\end{center}
\vspace*{-1.0cm} \caption{The temperature-number density phase
diagram of 2D $\rm ^3He$. The part for $z<1$ corresponds to HF
behavior divided into the LFL and NFL parts by the line
$T_M(z)\propto (1-z)^{3/2}$, where $T_M$ is the effective mass
maximum temperature. The exponent $3/2=1.5$ coming from Eq.
(\ref{zui4}) is in good agreement with the experimental value of
$1.7\pm 0.1$ \cite{he3}. The dependence $M^*(z)\propto (1-z)^{-1}$
shown by the dashed line points out QCP taking place at $z=1$. The
regime for $z\geq 1$ consists of the LFL piece (the shadowed
region, beginning in the intervening phase $z\leq 1$ \cite{he3},
which is due to the substrate inhomogeneities, see text) and the
NFL regime at higher temperatures.}\label{fd}
\end{figure}

In Fig. \ref{fd}, we show the phase diagram of 2D $^3$He in the
variables $T$ - $z$ (see Eq. \eqref{zui2}). For the sake of
comparison the plot of the effective mass versus $z$ is shown by
dashed line. The dependence $M^*(z)\propto (1-z)^{-1}$ demonstrates
that the effective mass diverges at QCP with $z=1$ in accordance
with the general phase diagram displayed in Fig. \ref{fig1}. The
part of the diagram where $z<1$ corresponds to HF behavior and
consists of LFL and NFL parts, divided by the line $T_M(z)\propto
(1-z)^{3/2}$. We draw attention here, that our exponents $1$ (see
Eq. \eqref{zui2}) and $3/2=1.5$ (see Eq. \eqref{zui4}) are in good
agreement with these from Ref. \cite{he3}. The good coincidence
between the theoretical and experimental exponents speaks in favor
of realization of our FCQPT scenario in the NFL behavior of both 2D
$^3$He and HF metals as former system is in great detail similar to
them.

The regime for $z>1$ located above the quantum critical line, see
Figs. \ref{fd} and \ref{fig1}, consists of low-temperature LFL
piece, (shown in Fig. \ref{fd} by shadowed region, beginning in the
intervening phase $z\leq 1$ \cite{he3}) and NFL regime at higher
temperatures. The former LFL piece is related to the peculiarities
of substrate on which 2D $\rm ^3He$ film is placed. Namely, it is
related to weak substrate heterogeneity (steps and edges on its
surface) so that quasiparticles, being localized (pinned) on it,
give rise to the LFL behavior \cite{he3,he3a}. That is the
peculiarities of the substrate eliminate the degeneracy generated
by the FC state taking place at $z>1$ in the same way as the AF
state does in the case of $\rm YbRh_2Si_2$, see Fig. \ref{PHD}. At
elevated temperatures, the competition between thermal and pinning
energies returns the system back to the unpinned state restoring
the NFL behavior. As HF metals do not have a substrate, the LFL
behavior is induced by the AF state lifting the degeneracy. At
elevated temperatures, this state is destroyed and exhibits the NFL
behavior, as it is shown in Fig. \ref{PHD}. If the AF state were
absent and some disorder (like point defects, dislocations etc)
were present in the lattice a rather thin LFL piece could take
place at low temperatures.

\subsection{NFL behavior of 2D $\rm \bf ^3He$ versus that of HF metals}

\begin{figure} [! ht]
\begin{center}
\includegraphics [width=0.4\textwidth]{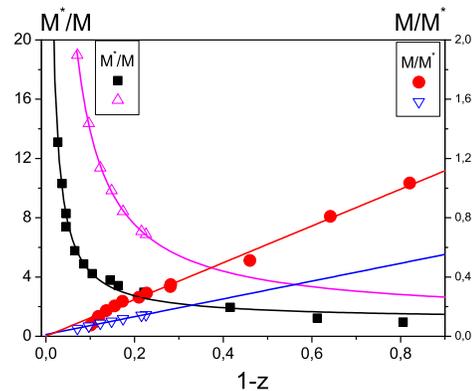}
\end{center}
\vspace*{-1.0cm} \caption{The dependence of the effective mass
$M^*(z)$ on the dimensionless density $1-z=1-x/x_c$. Experimental
data from Ref. \cite{prlhe} are shown by circles and squares and
those from Ref. \cite{he3} are shown by triangles. The effective
mass is fitted as $M^*(z)/M\propto A+B/(1-z)$ (see Eq.
(\ref{zui2})), while the reciprocal one as $M/M^*(z)\propto A_1z$,
where $A,B$ and $A_1$ are constants.}\label{MXM}
\end{figure}

As we have seen above, $M^*(T)$ can be measured in experiments on
strongly correlated Fermi-systems. For example, $M^*(T)\propto
C(T)/T\propto S(T)/T\propto M_0(T)\propto\chi(T)$ where $C(T)$ is
the specific heat, $S(T)$ --- entropy, $M_0(T)$ --- magnetization
and $\chi(T)$
--- AC magnetic susceptibility. If the measurements
are performed at fixed $x$ then, as it follows from Eq.
(\ref{UN2}), the effective mass reaches the maximum at $T=T_M$.
Upon normalizing both $M^*(T)$ by its peak value at each $x$ and
the temperature by $T_M$, we see from Eq. (\ref{UN2}) that in the
case of 2D $\rm ^3He$ all the curves also merge a into single one,
demonstrating a scaling behavior.

\begin{figure} [! ht]
\begin{center}
\includegraphics [width=0.4\textwidth]{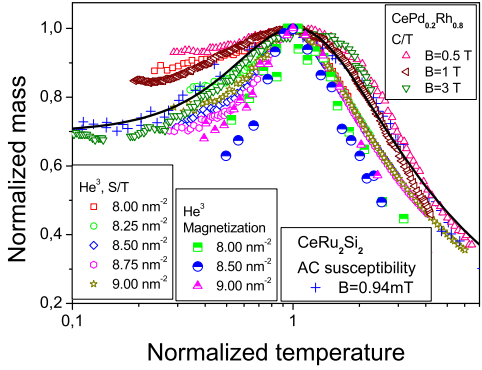}
\end{center}
\vspace*{-1.0cm} \caption{The normalized effective mass $M^*_N$ as
a function of the normalized temperature $T/T_M$ at densities shown
in the left bottom corner. The behavior of $M^*_N$ is extracted
from experimental data obtained in 2D $^3$He \cite{he3a} and 3D HF
compounds with different magnetic ground states such as
$\rm{CeRu_2Si_2}$ and $\rm CePd_{1-x}Rh_x$ \cite{pikul,takah},
fitted by the solid curve given by \eqref{UN2}.}\label{f2}
\end{figure}

In Fig. \ref{MXM}, we report the experimental values of effective
mass $M^*(z)$ obtained by the measurements on $^3$He monolayer
\cite{prlhe}. These measurements, in coincidence with those from
Ref. \cite{he3}, show the divergence of the effective mass at
$x=x_c$. To show that our FCQPT approach is able to describe the
above data, we represent the fit of $M^*(z)$ by the fractional
expression $M^*(z)/M\propto A+B/(1-z)$ and the reciprocal effective
mass by the linear fit $M/M^*(z)\propto A_1z$. We note here, that
the linear fit has been used to describe the experimental data for
a bilayer of $^3$He \cite{he3} and we use this function here for
the sake of illustration. It is seen from Fig. \ref{MXM} that the
data \cite{he3} ($^3$He bilayer) can be equally well approximated
by both linear and fractional functions, while the data
\cite{prlhe} cannot. For instance, both fitting functions give for
the critical density in bilayer $x_c\approx 9.8$ nm$^{-2}$, while
for monolayer \cite{prlhe} these values are different: $x_c=5.56$
for a linear fit and $x_c=5.15$ for a fractional fit. It is seen
from Fig. \ref{MXM}, that a linear fit is unable to properly
describe the experiment \cite{prlhe} at small $1-z$ (i.e. near
$x=x_c$), while the fractional fit describes the experiment very
well. This means that more detailed measurements are necessary in
the vicinity $x=x_c$.

\begin{figure} [! ht]
\begin{center}
\includegraphics [width=0.4\textwidth]{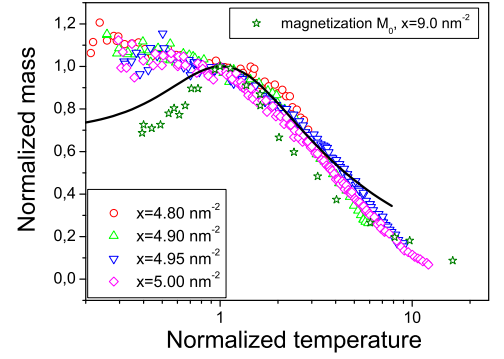}
\end{center}
\vspace*{-1.0cm} \caption{The dependence $M^*_N(T/T_M)$ at
densities shown in the left down corner. The behavior $M^*_N$ is
extracted from experimental data for $C(T)/T$ in 2D $^3$He
\cite{prlhe} and for the magnetization $M_0$ in 2D $^3$He
\cite{he3}. The solid curve shows the universal function, see the
caption to Fig. \ref{f2}.}\label{PRLC}
\end{figure}
We now apply the universal dependence \eqref{UN2} to fit the
experiment not only in 2D $^3$He but also in 3D HF metals.
$M^*_N(y)$ extracted from the entropy $S(T)/T$ and magnetization
$M_0$ measurements on the $^3$He film \cite{he3a} at different
densities $x<x_c$ is reported in Fig. \ref{f2}. In the same figure,
the data extracted from the heat capacity of ferromagnet
CePd$_{0.2}$Rh$_{0.8}$ \cite{pikul} and the AC magnetic
susceptibility of paramagnet CeRu$_2$Si$_2$ \cite{takah} are
plotted for different magnetic fields. It is seen that the
universal behavior of the normalized effective mass given by Eq.
(\ref{UN2}) and shown by the solid curve is in accord with the
experimental facts. All 2D $^3$He substances are located at
$x<x_c$, where the system progressively disrupts its LFL behavior
at elevated temperatures. In that case the control parameter,
driving the system towards its QCP $x_c$ is merely the number
density $x$. It is seen that the behavior of $M^*_N(y)$, extracted
from $S(T)/T$ and $M_0$ of 2D $^3$He (the entropy $S(T)$ is
reported in Fig. S8 A of Ref. \cite{he3a}) looks very much like
that of 3D HF compounds. As we shall see from Fig. \ref{STM} below,
the amplitude and positions of the maxima of magnetization $M_0(T)$
and $S(T)/T$ in 2D $^3$He follow nicely Eqs. \eqref{zui2} and
\eqref{zui4}. We conclude that Eq. \eqref{UN2} allows for the
reduction of a 4D function describing the effective mass to a
function of a single variable. Indeed, the effective mass depends
on the magnetic field, temperature, number density and composition
so that all these parameters can be merged in the single variable
by means of interpolating function like Eq. \eqref{UN2}.

The attempt to fit the available experimental data for $C(T)/T$ in
2D $\rm ^3He$ \cite{prlhe} by the universal function $M^*_N(y)$ is
reported in Fig. \ref{PRLC}. Here, the data extracted from heat
capacity $C(T)/T$ for the $^3$He monolayer \cite{prlhe} and
magnetization $M_0$ for the bilayer \cite{he3}, are reported. It is
seen that the effective mass extracted from these thermodynamic
quantities can be well described by the universal interpolation
formula \eqref{UN2}. We note the qualitative and quantitative
similarity between the double layer \cite{he3} and the monolayer
\cite{prlhe} of $^3$He as seen from Fig. \ref{PRLC}.

\begin{figure} [! ht]
\begin{center}
\includegraphics [width=0.4\textwidth]{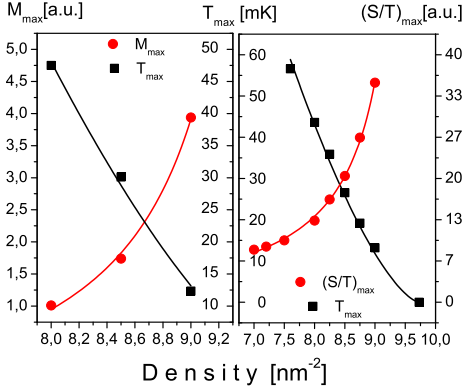}
\end{center}
\vspace*{-1.0cm} \caption{Left panel, the peak temperatures $T_M$
and the peak values $M_{\rm max}$ extracted from measurements of
the magnetization $M_0$ in $^3$He \cite{he3}. Right panel shows
$T_M$ and the peak values $(S/T)_{\rm max}$ extracted from
measurements of $S(T)/T$ in $^3$He \cite{he3a}. We approximate
$T_M\propto (1-z)^{3/2}$ and $(S/T)_{\rm max}\propto M_{\rm
max}\propto A/(1-z)$.}\label{STM}
\end{figure}

In the left panel of Fig. \ref{STM}, we show the density dependence
of $T_M$, extracted from measurements of the magnetization $M_0(T)$
of the $^3$He bilayer \cite{he3}. The peak temperature is fitted by
Eq. \eqref{zui4}. In the same Figure, we have also reported the
maximal magnetization $M_{\rm max}$. It is seen that $M_{\rm max}$
is well described by the expression $M_{\rm max} \propto (S/T)_{\rm
max}\propto (1-z)^{-1}$, see Eq. (\ref{zui2}). The right panel of
Fig. \ref{STM} reports the peak temperature $T_M$ and the maximal
entropy $(S/T)_{\rm max}$ versus the number density $x$. They are
extracted from the measurements of $S(T)/T$ on the $^3$He bilayer
\cite{he3a}. The fact that both the left and right panels exhibit
the same behavior of the curves shows once more that there are
indeed the quasiparticles, determining the thermodynamic behavior
of 2D $^3$He near its QCP related to FCQPT.

\section{Summary}

We have analyzed the non-Fermi liquid behavior of the heavy fermion
metals, and showed that extended quasiparticles paradigm is
strongly valid, while the dependence of the effective mass on
temperature, number density and applied magnetic fields gives rise
to the NFL behavior. We have demonstrated that our theoretical
study of the heat capacity, magnetization, longitudinal
magnetoresistance and magnetic entropy are in good agreement with
the outstanding recent facts collected on the HF metal $\rm
YbRh_2Si_2$. Our normalization procedure has allowed us to
construct the scaled thermodynamic and transport properties in wide
range of the variation of the scaled variable. For $\rm YbRh_2Si_2$
the constructed thermodynamic and transport functions show the
scaling behavior over three decades in the normalized variable. The
energy scales in these functions are also explained.

We have described the diverse experimental facts related to
temperature and number density dependencies of the thermodynamic
characteristics of 2D $^3$He by a single universal function of one
argument. The above universal behavior is also inherent to HF
metals with different magnetic ground states. We obtain the
marvelous coincidence with experiment in the framework of our
theory. Moreover, these data could be obtained for 2D $^3$He only
and thus they were inaccessible for analysis in HF metals. This
fact also shows the universality of our approach. Thus we have
shown that bringing the experimental data collected on different
strongly correlated Fermi-systems to the above form immediately
reveals their universal scaling behavior. Thus, the theory of
fermion condensation quantum phase transition, preserving the
extended quasiparticles paradigm and intimately related to
unlimited growth of the effective mass as a function of
temperature, magnetic field etc, is capable of describing the
strongly correlated Fermi systems.

\section{Acknowledgement}

This work was supported in part by RFBR No. 09-02-00056.

\end{document}